\documentclass[aps,PRB,twocolumn,english,superscriptaddress,citeautoscript,preprintnumbers,amsmath,amssymb,floatfix,footinbib]{revtex4-1}
\usepackage{graphicx}
\usepackage{graphics}               
\usepackage{subfigure}
\usepackage{bm}   
\usepackage{amssymb } 
\usepackage{array}       
\usepackage[breaklinks=true,colorlinks,citecolor=blue,linkcolor=blue,urlcolor=blue]{hyperref}
\usepackage{float}
\usepackage[table]{xcolor}
\usepackage{cancel}
\usepackage[normalem]{ulem}

\definecolor{Cyan}{rgb}{0.4,1,0.7}
\definecolor{LightCyan}{rgb}{0.8,1,0.85}

\begin{document}
	
\title{Nernst power factor and figure of merit in compensated semimetal ScSb}

\author{Antu Laha}\email[]{antulaha.physics@gmail.com}
    \affiliation{Department of Physics and Astronomy, Stony Brook University, Stony Brook, New York 11794-3800, USA}
    \affiliation{Condensed Matter Physics and Materials Science Division, Brookhaven National Laboratory, Upton, New York 11973-5000, USA}
 
 \author{Sarah Paone}
    \affiliation{Department of Physics and Astronomy, Stony Brook University, Stony Brook, New York 11794-3800, USA}
    \affiliation{Condensed Matter Physics and Materials Science Division, Brookhaven National Laboratory, Upton, New York 11973-5000, USA}

\author{Asish K. Kundu }
     \affiliation{National Synchrotron Light Source II, Brookhaven National Laboratory, Upton, New York 11973-5000, USA}
    
\author{Juntao Yao}
    \affiliation{Condensed Matter Physics and Materials Science Division, Brookhaven National Laboratory, Upton, New York 11973-5000, USA}
    \affiliation{Department of Materials Science and Chemical Engineering, Stony Brook University, Stony Brook, New York 11794-3800, USA}

\author{Niraj Aryal }
    \affiliation{Condensed Matter Physics and Materials Science Division, Brookhaven National Laboratory, Upton, New York 11973-5000, USA}    
\author{Elio Vescovo }
     \affiliation{National Synchrotron Light Source II, Brookhaven National Laboratory, Upton, New York 11973-5000, USA}

\author{Qiang Li}\email[]{qiang.li@stonybrook.edu}
    \affiliation{Department of Physics and Astronomy, Stony Brook University, Stony Brook, New York 11794-3800, USA}
     \affiliation{Condensed Matter Physics and Materials Science Division, Brookhaven National Laboratory, Upton, New York 11973-5000, USA}

\begin{abstract}

Recently, topological semimetals have emerged as strong candidates for solid-state thermomagnetic refrigerators due to their enhanced Nernst effect. This enhancement arises from the combined contributions of the Berry-curvature-induced anomalous Nernst coefficient associated with topological bands and the normal Nernst effect resulting from synergistic electron-hole compensation. Generally, these two effects are intertwined in topological semimetals, making it challenging to evaluate them independently. Here, we report the observation of high Nernst effect in the electron–hole compensated semimetal ScSb with topologically trivial electronic band structures. Remarkably, we find a high maximum Nernst power factor of $PF_N \sim 35 \times 10^{-4}$ W m$^{-1}$ K$^{-2}$ in ScSb. The Nernst thermopower ($S_{xy}$) exhibits a peak of $\sim$ 47 $\mu$V/K at 12 K and 14 T, yielding a Nernst figure of merit ($z_N$) of $\sim 28 \times 10^{-4}$ K$^{-1}$. Notably, despite its trivial electronic band structure, both the $PF_N$ and $z_N$ values of ScSb are comparable to those observed in topological semimetals with Dirac band dispersions. The origin of the large Nernst signal in ScSb is explained by well compensated electron and hole carriers, through Hall resistivity measurements, angle-resolved photoemission spectroscopy (ARPES) and density functional theory (DFT) calculations.

\end{abstract}

\maketitle

\section{Introduction}
Low-temperature environments are essential for many areas of scientific research, including the study of quantum states of matter, superconducting materials, and space science. Traditionally, refrigeration systems that use cryogenic liquids and gases have been the most common way to achieve these low temperatures. However, thermoelectric (TE) refrigeration presents an attractive alternative, offering the ability to generate cooling through solid-state devices without the need for moving parts. The principle behind TE refrigeration is based on the Seebeck-Peltier effect, where an electric current is applied to a thermoelectric material, causing one side to become cold while the other side heats up. This method is fast and precise, but the cooling power of Peltier devices have some limitations. Most TE devices today are made by connecting n-type and p-type materials in a series arrangement. The contact points between these materials create resistance to both electricity and heat, which reduces the overall efficiency of the device. Another type of refrigeration method based on solid-state devices, Ettingshausen refrigeration, has not been as extensively explored as Peltier refrigeration but holds significant potential for cooling technologies. Ettingshausen refrigeration requires only a single material, unlike Peltier devices that need coupled p-type and n-type legs, making the device construction much simpler. The working principle of the Ettingshausen refrigerator is based on the Nernst–Ettingshausen effect, where an electric current passing through a material exposed to a magnetic field generates a temperature gradient that is perpendicular to both the current and the magnetic field.

Interest in Nernst–Ettingshausen effect has been renewed in light of significant advancements in topological semimetals. Several of these materials such as Cd$_3$As$_2$, ZrTe$_5$, NbP, PtSn$_4$, Mg$_2$Pb, NbSb$_2$, and WTe$_2$ have been identified as exhibiting strong thermomagnetic performance \cite{Cd3As2_PRL_2017, Cd3As2_SCP_2020, ZrTe5_PRB_2021, NbP_JPCM_2017, NbP_PRB_2016, PtSn4_Research_2020, Mg2Pb_natComn_2021, NbSb2_NatComn_2022, WTe2_NanoLett_2018, WTe2_NatComn_2022}. Three factors are claimed to act synergistically in enhancing the Nernst thermopower $S_{xy}$: (i) phonon drag effect, (ii) electron-hole compensation effect, and (iii) topologically non-trivial electronic bands. An important question that arises is whether non-trivial topology is essential for observing a strong Nernst effect. To address this, we investigate the Nernst effect in a topologically trivial semimetal ScSb, which simultaneously hosts perfect electron-hole compensation and the phonon drag effect \cite{ScSb_PRB_2018, ScSb_PRB_2021}. Interestingly, ScSb shows a strong Nernst peak of $\sim$ 47 $\mu$V/K at 14 T and 12 K. This leads to a maximum Nernst power factor ($PF_N$) of $\sim$ 35 $\times 10^{-4}$ W m$^{-1}$ K$^{-2}$ and Nernst figure of merit ($z_N$) of $\sim 28 \times 10^{-4}$ K$^{-1}$. The Nernst power factor for ScSb is comparable to the Seebeck power factor ($PF_S$) found in state-of-the-art thermoelectric materials such as Bi$_2$Te$_3$ ($PF_S$ = 42 $\times 10^{-4}$ W m$^{-1}$ K$^{-2}$), PbTe ($PF_S$ = 25 $\times 10^{-4}$ W m$^{-1}$ K$^{-2}$), and SnSe ($PF_S$ = 10 $\times 10^{-4}$ W m$^{-1}$ K$^{-2}$). Notably, despite possessing a trivial electronic band structure, both the $PF_N$ and $z_N$ values of ScSb are comparable to those observed in topological semimetals with Dirac band dispersions, such as Cd$_3$As$_2$, ZrTe$_5$, PtSn$_4$, and TbPtBi \cite{Cd3As2_PRL_2017, Cd3As2_SCP_2020, ZrTe5_PRB_2021, PtSn4_Research_2020, TbPtBi_AM_2023}. The pronounced Nernst signal in ScSb is attributed to well compensated electron and hole carriers, via Hall resistivity measurements, angle-resolved photoemission spectroscopy (ARPES), and density functional theory (DFT) calculations.

\section{Methods}

\textbf{Synthesis of single crystal.}
ScSb single crystals were grown by the standard self flux method with excess Sb as a flux \cite{ScSb_PRB_2018, ScSb_PRB_2021}. Sc ingot (99.99$\%$), and Sb chips (99.999$\%$) in a molar ratio of 1:15 were mixed in an alumina crucible. The crucible was then sealed into a quartz tube under a partial pressure of argon gas. The content was heated to 1100$^\circ$C, kept for 12 hours at that temperature, and then cooled to 750$^\circ$C at a rate of 2$^\circ$C/hour. Cubic shaped single crystals were extracted from the flux by centrifuging. The typical size of the crystal is 2.5 mm $\times$ 2.5 mm $\times$ 2 mm.

\begin{figure}
	\centering
	\includegraphics[width=0.99\linewidth]{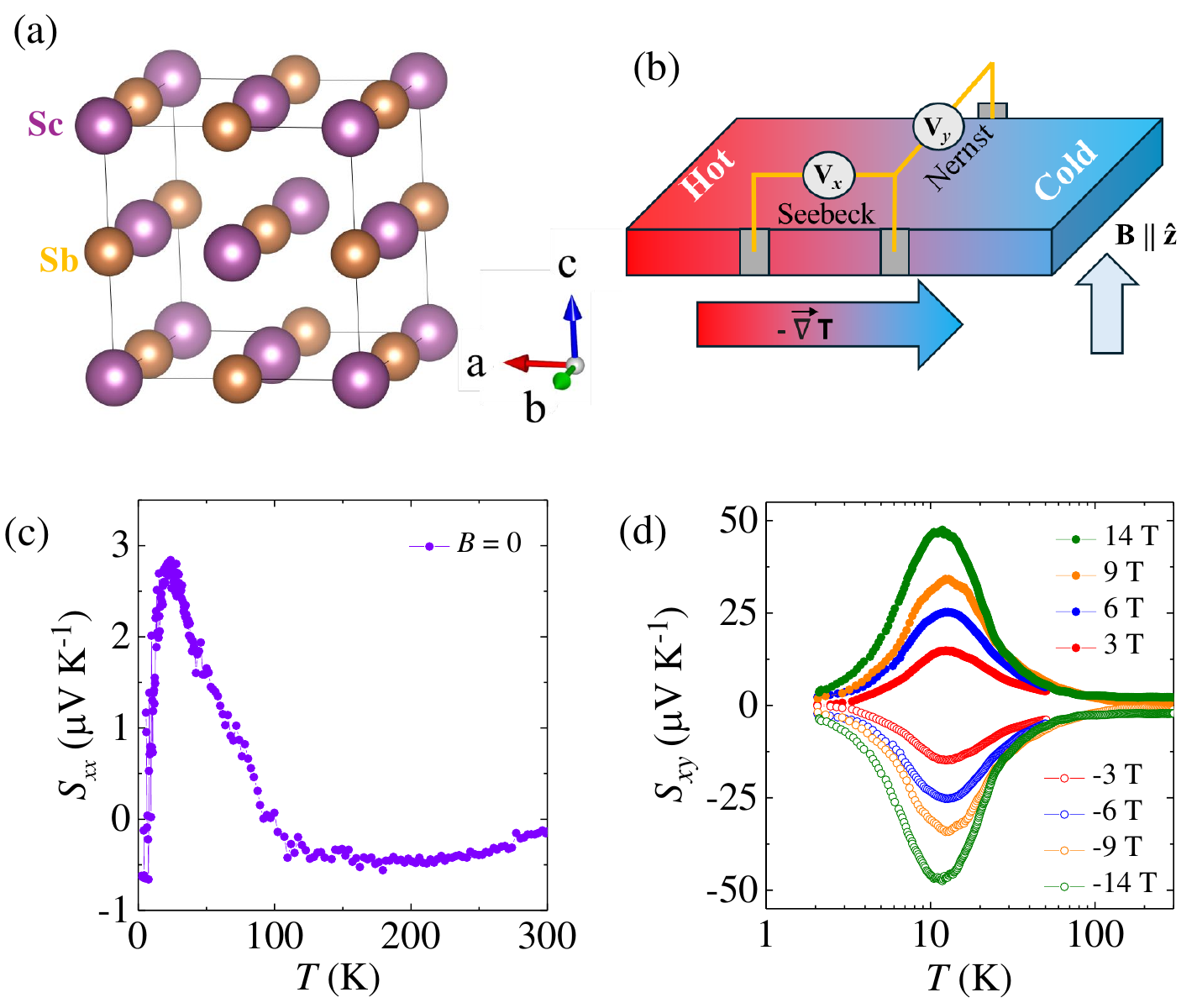} 
	\caption{(a) Cubic crystal structure of ScSb. (b) Schematic diagram for Seebeck and Nernst effect measurements. (c) Seebeck thermopower ($S_{xx}$), and (d) Nernst thermopower ($S_{xy}$) as a function of temperature.}
	\label{Fig1}
\end{figure}

\textbf{Characterization and transport property measurements.}
The powders of crushed single crystals were characterized by x-ray
diffraction (XRD) technique using Cu-K$_\alpha$ radiation in a Rigaku miniflex diffractometer to determine the crystal structure and the phase purity. The XRD data were analyzed by Rietveld structural refinement using the FULLPROF package [see Supplemental Material (SM)]\cite{Fullprof_1993}. The chemical compositions were confirmed by energy dispersive x-ray spectroscopy (EDS) measurements in a a JEOL JSM-7600F scanning electron microscope [see SM]. Electronic transport measurements were carried out in a physical property measurement system (PPMS, Quantum Design) via the standard four-probe method. The Seebeck thermopower and Nernst thermopower were measured using four-probe technique on a standard thermal transport option (TTO) platform, and a modified one where the Cu wires for measuring voltage signals were separated from the Cernox 1050 thermometers \cite {NbSb2_NatComn_2022}.

\textbf{First principles calculations.}
The \textit{ab-initio} calculations for ScSb were done using the Quantum Espresso~\cite{giannozzi2009quantum} implementation of the density functional theory (DFT) in the Generalized Gradient Approximation (GGA) framework including spin-orbit coupling. Fully relativistic norm-conserving pseudopotentials and  Perdew-Burke-Ernzerhof (PBE) exhange-correlation functional were used~\cite{perdew1996generalized}. Convergence testing for the kinetic energy cutoff and \textit{k-mesh} was done to achieve total energy convergence of less than 1 meV which led  to the energy cutoff value of 300 Ry and a \textit{k-mesh} grid of 20 $\times$ 20 $\times$ 20. Finally, the fermi surface was calculated by using a dense k mesh-grid of 30 $\times$ 30 $\times$ 30.\\

\textbf{ARPES measurements.}
ARPES experiments were performed at the Electron Spectro Microscopy (ESM) 21-ID-1 beamline of the National Synchrotron Light Source II at Brookhaven National Lab \cite{rajapitamahuni2024electron}. The beamline is equipped with a Scienta DA30 electron analyzer, with base pressure $\sim$ 1$\times$10$^{-11}$ mbar. Prior to the ARPES experiments, samples were cleaved using a post inside an ultra-high vacuum chamber (UHV) at $\sim$ 18 K. The total energy and angular resolution during the ARPES measurements were $\sim$ 10 meV and $\sim$ 0.2$^\circ$, respectively.

\begin{figure}
	\centering
	\includegraphics[width=0.99\linewidth]{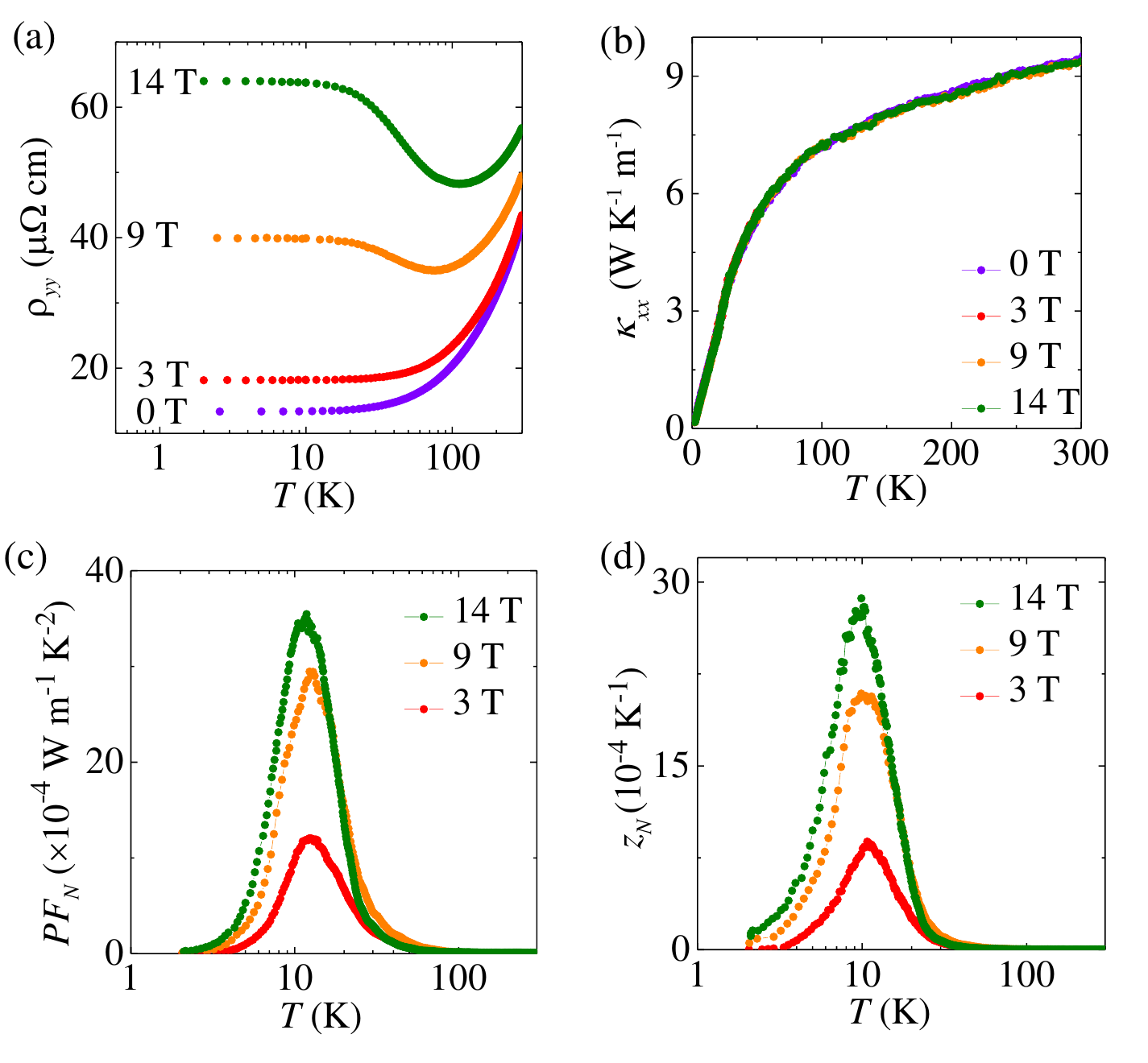} 
	\caption{(a) Longitudinal electrical resistivity ($\rho_{yy}$), (b) Longitudinal thermal conductivity ($\kappa_{xx}$), (c) Nernst power factor ($PF_N$), and (d) Nernst figure of merit ($z_N$) as a function of temperature.}
	\label{Fig2}
\end{figure}
\section{Results and Discussions}
Seebeck thermopower ($S_{xx}$) and Nernst thermopower ($S_{xy}$) are measured as a function of temperature for ScSb single crystal as shown in Fig.\ref{Fig1}(c) and \ref{Fig1}(d). The value of $S_{xx}$ is nearly zero down to $T$ = 100 K, suggesting electron-hole compensation. Below 100 K, the $S_{xx}$ increases with decreasing temperature and shows a strong peak at $\sim$ 12 K. A pronounced peak in $S_{xx}$ has been reported in various thermoelectric and thermomagnetic materials, commonly attributed to the phonon-drag effect. As temperature rises, phonons with greater momentum become more active. When the momentum of long-wavelength acoustic phonons aligns with that of charge carriers near the Fermi surface, the phonon-drag effect emerges, manifesting as a low temperature peak in the Seebeck thermopower \cite{Goldsmid, Delves_1964}. However, at higher temperatures, this effect diminishes rapidly due to increased excitation of high-frequency phonons, which in turn reduce the relaxation time of the acoustic phonons. The same mechanism also boosts the Nernst thermopower, evidenced by a peak observed around 12 K [Fig.\ref{Fig1} (d)]. The peak in $S_{xy}$ becomes more pronounced at higher magnetic fields, and it reaches 47 $\mu$V/K at 14 T. The $S_{xy}$ shows similar behavior with negative sign under the application of -14 T magnetic fields. Such a large enhancement of $S_{xy}$ previously observed in several topological semimetals is generally contributed to the synergistic effect of non-trivial electronic bands and electron-hole compensation effect. Here, we observe the enhancement of Nernst thermopower in a compensated semimetal ScSb having only topologically trivial bands. \\

\begin{figure}
	\centering
	\includegraphics[width=1.0\linewidth]{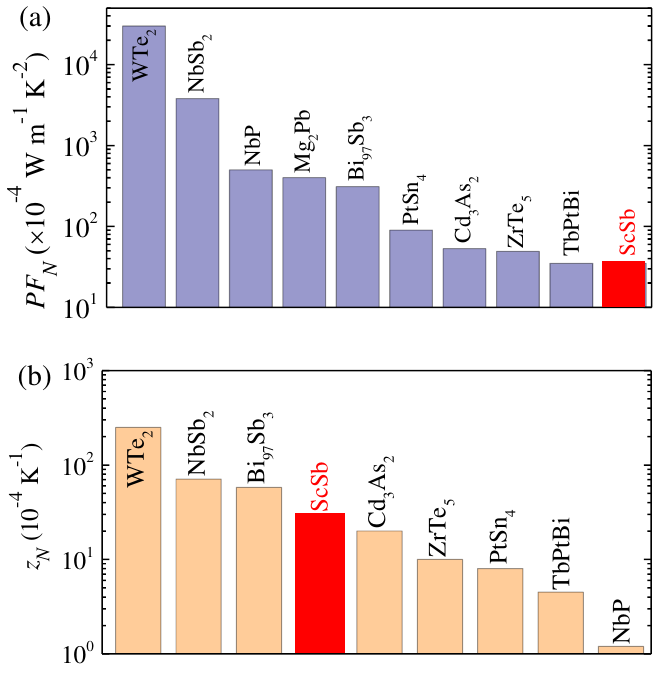} 
	\caption{Comparison of the (a) Nernst power factor $PF_N$ and (b) Nernst figure of merit $z_N$ of ScSb single crystal and some typical thermomagnetic topological semimetals \cite{WTe2_NatComn_2022, NbSb2_NatComn_2022, NbP_JPCM_2017, NbP_PRB_2016, Mg2Pb_natComn_2021, PtSn4_Research_2020, Cd3As2_PRL_2017, ZrTe5_PRB_2021, TbPtBi_AM_2023, Bi97Sb3_PRB_2018}.}
	\label{Fig_cmp}
\end{figure}
\begin{figure}
	\centering
	\includegraphics[width=0.99\linewidth]{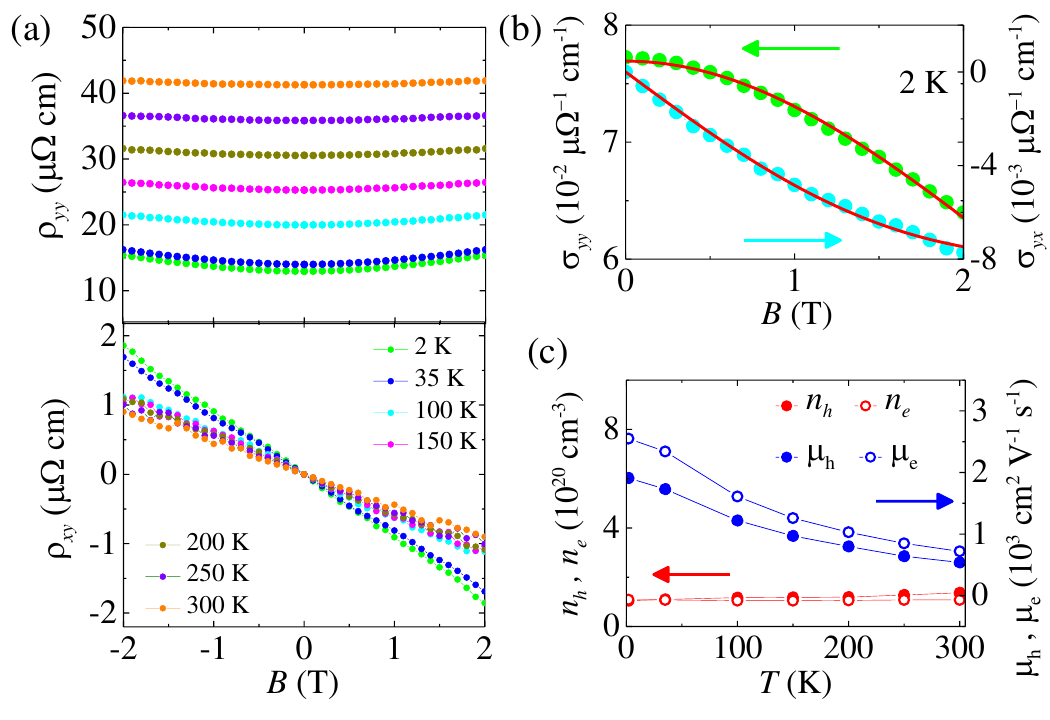} 
	\caption{(a) Longitudinal resistivity ($\rho_{yy}$) and Hall resistivity ($\rho_{xy}$) as function of $B$ at various temperatures. (b) Simultaneous fitting of longitudinal conductivity ($\sigma_{yy}$) and Hall conductivity ($\sigma_{yx}$) with two carries model (equation (\ref{yy}) and (\ref{yx})). (c) Hole (electron) density $n_h$ ($n_e$) and mobility $\mu_h$ ($\mu_e$) as a function of temperature.}
	\label{Fig_Hall}
\end{figure}
\begin{figure*}[htp]
	\centering
	\includegraphics[width=1\linewidth]{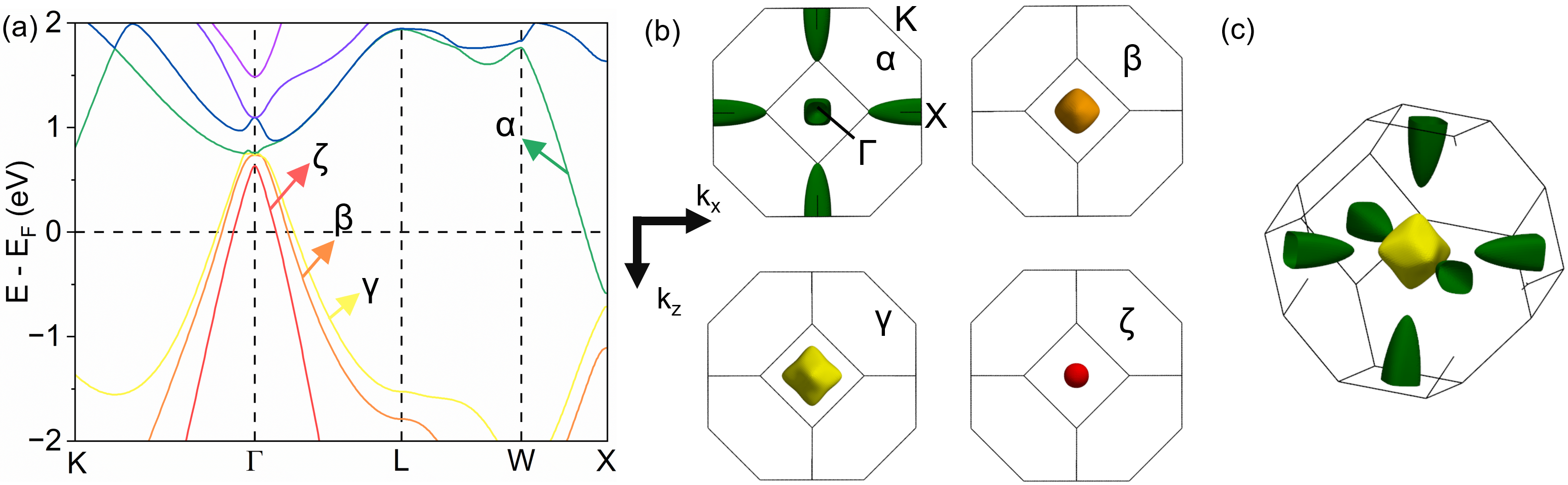} 
	\caption{ Electronic structure of ScSb. (a) Band structure of along a labeled high-symmetry path. Labels for the Fermi pockets are shown on the band structure. (b) Individual Fermi pockets projected on $k_x-k_z$ plane with high symmetry points labeled on the top left Brillouin zone. Green corresponds to the electron side pockets, while red, orange, and yellow show the $\Gamma$-centered nested hole pockets. (c) 3-dimensional view of the Fermi surface showing ellipsoidal electron pockets  around the X-point (zone boundary) and $\Gamma$-centered largest hole pocket ($\gamma$ pocket ) which envelopes $\beta$ and $\zeta$ hole pockets.}
	\label{ScSb_band_fermi}
\end{figure*}
To determine the Nernst power factor ($PF_N$) and Nernst figure of merit ($z_N$), we measure the longitudinal electrical resistivity ($\rho_{yy}$) and longitudinal thermal conductivity ($\kappa_{xx}$). The temperature dependence of $\rho_{yy}$ displays a metal-semiconductor-like crossover followed by a resistivity plateau in presence of magnetic field, which becomes more noticeable as $B$ is increased from 3 T to 14 T [see Fig.\ref{Fig2}(a)]. The field induced  metal-semiconductor-like crossover and resistivity plateau are generic features of electron-hole compensated semimetals \cite{ScSb_PRB_2018, LaSb_NaturePhy_2015, TmSb_PRB_2018, LaSbBi_PRB_2018, CdCdSn_PRB_2020, YbCdSn_PRB_2020, YbCdGe_PRB_2019}. The temperature dependence of longitudinal thermal conductivity ($\kappa_{xx}$) is shown in Fig.\ref{Fig2}(b). The $\kappa_{xx}$ decreases with decreasing temperature and reaches $\sim$ 0.2 W K$^{-1}$ m$^{-1}$ at 2 K. At room temperature, the thermal conductivity of ScSb ($\sim 10 $ W K$^{-1}$ m$^{-1}$) is higher than that of some well-known thermoelectric materials such as Bi$_2$Te$_3$ ($\sim$ 1.2 W K$^{-1}$ m$^{-1}$), PbTe ($\sim$ 2.3 W K$^{-1}$ m$^{-1}$), Mg$_3$Bi$_2$ ($\sim$ 2 W K$^{-1}$ m$^{-1}$) and filled skutterudites ($\sim$ 3.0 W K$^{-1}$ m$^{-1}$) \cite{Mg3Bi2_Science_2019, Bi2Te3_kxx_2002, Skutterudites_JACS_2011}. However, the $\kappa_{xx}$ is only $\sim$ 1.5 W K$^{-1}$ m$^{-1}$ at 12 K where ScSb shows a Nernst peak. This is significantly low compared to that of the topological semimetals exhibiting Nernst peak, such as NbP (1290 W K$^{-1}$ m$^{-1}$) \cite{NbP_JPCM_2017}, TaP (1586 W K$^{-1}$ m$^{-1}$) \cite{TaP_NatComn_2020}, WTe$_2$ (215 W K$^{-1}$ m$^{-1}$) \cite{WTe2_NatComn_2022}, and NbSb$_2$ (90 W K$^{-1}$ m$^{-1}$) \cite{NbSb2_NatComn_2022}. The  Nernst power factor and Nernst figure of merit are calculated using the relations $PF_N = S_{xy}^2 \sigma_{yy}$ and $z_N = S_{xy}^2 \sigma_{yy}/\kappa_{xx}$ as displayed in Fig.\ref{Fig2}(c) and \ref{Fig2}(d). We note that longitudinal transport efficiencies ($\kappa$ and $\rho$) along $x$ and $y$ direction should be the same for ScSb has a cubic crystalline structure. The $PF_N$ of ScSb reaches a peak of $\sim 35 \times 10^{-4}$ W m$^{-1}$ K$^{-2}$ at 12 K and 14 T. The $PF_N$ for ScSb is comparable to the Seebeck power factor ($PF_S$) found in state-of-the-art thermoelectric materials such as Bi$_2$Te$_3$ ($PF_S$ = 42 $\times 10^{-4}$ W m$^{-1}$ K$^{-2}$) \cite{Bi2Te3_Science_2008}, CoSb$_3$ ($PF_S$ = 40 $\times 10^{-4}$ W m$^{-1}$ K$^{-2}$) \cite{CoSb3_NatMatr_2015}, PbTe ($PF_S$ = 25 $\times 10^{-4}$ W m$^{-1}$ K$^{-2}$) \cite{PbTe_Nature_2011}, and SnSe ($PF_S$ = 10 $\times 10^{-4}$ W m$^{-1}$ K$^{-2}$) \cite{SnSe_Nature_2014}. Notably, despite trivial electronic band structure, ScSb has the maximum $PF_N$ value comparable to topological semimetals Cd$_3$As$_2$, ZrTe$_5$, PtSn$_4$, and TbPtBi \cite{Cd3As2_PRL_2017, Cd3As2_SCP_2020, ZrTe5_PRB_2021, PtSn4_Research_2020, TbPtBi_AM_2023}, while it has a larger $z_N$ (28 $\times 10^{-4}$ K$^{-1}$) owing to lower thermal conductivity. The maximum values of $PF_N$ and $z_N$ for several topological thermomagnetic materials, including ScSb, are shown in Fig.\ref{Fig_cmp}. \\

\begin{figure*}
	\centering
	\includegraphics[width=0.95\linewidth]{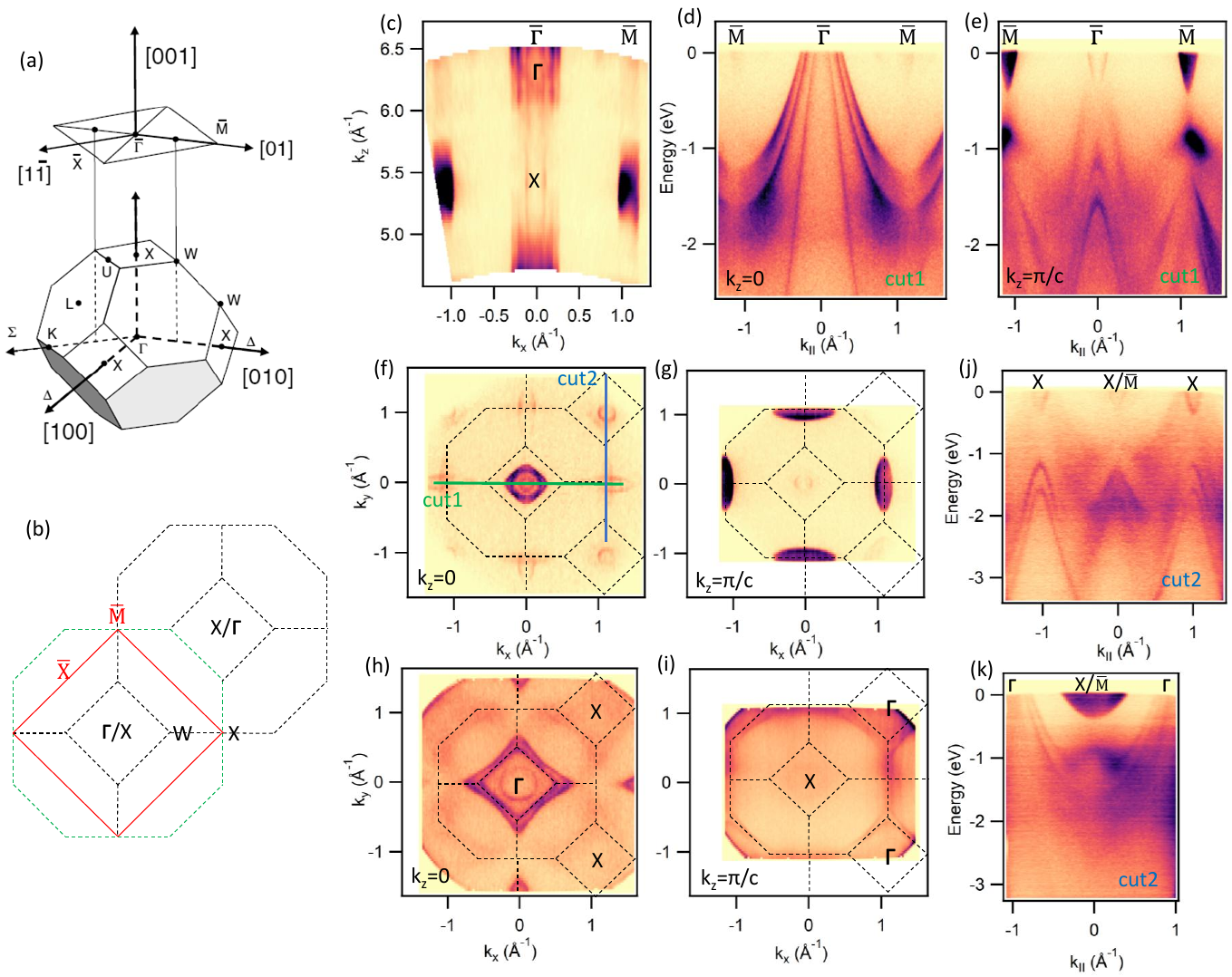} 
	\caption{Electronic structure of ScSb probed using ARPES at 20 K. (a) Primitive 3D Brillouin zone (BZ) of ScSb (bottom) and its surface projection [SBZ(top)]. (b) Projected 3D BZ on [001] plane and SBZ (red square). Two consecutive BZ are shown, where green dashed lines outline the boundary of first BZ at $k_z=0$ and the square formed by the black dashed lines inside it appears from $k_z=\pi/c$ plane. (c) Out-of-plane Fermi surface (FS) map using a wide range of photon energies (h$\nu$ = 80--160 eV). (d and e) Cut at $k_z=0$ ($\Gamma$) and $k_z=\pi/c$ (X) planes, respectively. (f and g) FS maps at $k_z=0$ and $k_z=\pi/c$ (w.r.t the first BZ), respectively. (h and i) Constant energy contours at -1 eV  for $k_z=0$ and $k_z=\pi/c$, respectively. (j and k) Band dispersion along cut2 as shown in (f) for $k_z=0$ and $k_z=\pi/c$, respectively.}
	\label{Fig_ARPES}
\end{figure*}

To determine the charge carriers densities and mobilities of ScSb, we measured magnetic field dependence of Hall resistivity ($\rho_{xy}$) and longitudinal resistivity ($\rho_{yy}$) at various temperatures from 2 K to 300 K [see Fig.\ref{Fig_Hall}(a)]. The field dependence of $\rho_{xy}$ is not linear, which indicates that both electrons and holes contribute to the transport. The Hall conductivity ($\sigma_{yx}$) and longitudinal conductivity ($\sigma_{yy}$) are simultaneously fitted with the semiclassical two carriers model \cite{Two_Carrier_Model} as shown in Fig.\ref{Fig_Hall}(b), 

\begin{equation}
  \sigma_{yy} = e \left[\frac{n_h \mu_h}{1 + \mu_h^2 B^2} + \frac{n_e \mu_e}{1 + \mu_e^2 B^2} \right] 
  \label{yy}
\end{equation}

\begin{equation}
     \sigma_{yx} = e B \left[\frac{n_h \mu_h^2}{1 + \mu_h^2 B^2} - \frac{n_e \mu_e^2}{1 + \mu_e^2 B^2}\right]
     \label{yx}
\end{equation}
where $n_h$ ($n_e$) and $\mu_h$ ($\mu_e$) are the hole (electron) density and mobility, respectively. The $\sigma_{yy}$ and $\sigma_{yx}$ are obtained using the expressions $\sigma_{yy} = \frac{\rho_{yy}}{\rho_{yy}^2 + \rho_{xy}^2}$, and $\sigma_{yx} = \frac{\rho_{xy}}{\rho_{yy}^2 + \rho_{xy}^2}$. The calculated carrier densities  (error bar: $\pm$5$\%$) and mobilities (error bar: $\pm 5\%$) from the two carriers model fitting are plotted as a function of temperature in Fig.\ref{Fig_Hall}(c). The $n_h/n_e$ ratio is $\sim 1.05\pm 5\%$, which confirms the nearly perfect electron-hole compensation in this compound.\\

To get further understanding of the electronic structure of ScSb, we performed density functional theory calculations. The band structure plot along a high symmetry path is shown in Fig.~\ref{ScSb_band_fermi}(a). The electronic structure reveals three nested hole pockets at the $\Gamma$-point in the Brillouin zone, accompanied by an electron side band at the X-point. The 2D and 3D view of the corresponding Fermi surface  are  shown in Figs.~\ref{ScSb_band_fermi}(b) and (c), respectively. The electron bands, labeled as $\alpha$, manifest as 3 pairs of semi-ellipsoids at the square faces of the Brillouin zone. We can also see the three hole bands labeled as $\gamma$, $\beta$, and $\zeta$ going from the largest to the smallest size centered around the $\Gamma$-point. No band inversion is found in ScSb, consistent with its classification as a topologically trivial semimetal. The charge carrier compensation extracted from experiment can be compared to the compensation calculated from the fermi surface volume by the DFT, where the ratio between electrons and holes is proportional to the volume for their respective pockets. The $n_h\slash n_e$ ratio was found to be $\sim 1.31$ by calculating the ratio of the total volumes of the hole to the electron pockets. The experimentally determined compensation, on the other hand, was found to be close to 1, within 5$\%$ error bar,  by fitting a semiclassical two carrier model to the Hall conductivity.  It is noted that Hu \textit{et al.} also estimated $n_e\slash n_p$ ratio of 0.93 using Shubnikov-de Haas oscillations \cite{ScSb_PRB_2018}.  
What this discrepancy between theory and experiment may indicate is that the Fermi level in the actual sample may be around 0.04 eV higher than the predicted Fermi level from the DFT shown in Fig.~\ref{ScSb_band_fermi}(a). 

Further we performed ARPES measurements to experimentally probe the electronic structure of ScSb, as presented in Fig.\ref{Fig_ARPES}. The schematic of three dimensional primitive Brillouin zone (BZ) and its projection on the [001] plane are shown in Figs. \ref{Fig_ARPES}(a) and (b), respectively. The projection of two consecutive BZ are shown for better understanding of the symmetry point location when compared to the ARPES results. Fig. \ref{Fig_ARPES}(c) represents the out-of-plane Fermi surface (FS) map using a wide range of photon energies ($h\nu$ = 80--160 eV). The $\Gamma$ and X points are identified at $k_z=6\times\pi/c$ and $k_z=5\times\pi/c$ (where, $\pi/c=1.067$ \AA$^{-1}$), which we consistently refer to as $k_z=0$ and $k_z=\pi/c$ throughout the text for simplicity. Around the X point, a single Fermi surface (FS) is observed, whereas around $\Gamma$, three distinct FS are evident. The dispersion of these states, as depicted in Figs. \ref{Fig_ARPES}(d) and (e), indicates that they originate from three hole pockets centered at $\Gamma$ and an electron pocket centered at X points, respectively. An electron pocket was also observed around $\bar{M}$, this is due to the fact that the X point is basically projected to the $\bar{M}$ point due to the cubic symmetry of the crystal (Fig. \ref{Fig_ARPES}(b)). 

The in-plane FS maps and constant energy cuts at $k_z=0$ and $k_z=\pi/c$ are shown in Figs. \ref{Fig_ARPES}(f-g) and Figs. \ref{Fig_ARPES}(h-i), respectively. From these plots its evident that the cross-section of the inner two hole pockets at $\Gamma$ are mostly circular, while the outermost hole pocket is square. On the other hand, the electron pockets at X point are elliptical. The dispersion of the electronic states along cut2 from Figs. \ref{Fig_ARPES}(f) and (g) are shown in Figs. \ref{Fig_ARPES}(j) and (k), respectively. The dispersion of the electron pocket at X shows strongly anisotropic nature.  Moreover, we identified three hole pockets at the BZ center and an electron pocket at six equivalent X-points (total 3 pairs of electron pocket). All these Fermi surfaces and their shape are in good agreement with our theoretical results. Further, we extracted the $k_\mathrm{F}$ values of the hole pockets (inner to outer: $\zeta$, $\beta$, and $\gamma$) along $\bar{\Gamma}$-$\bar{M}$ and $\bar{\Gamma}$-$\bar{X}$: the obtained values are 0.1, 0.2, 0.27 \AA$^{-1}$ and 0.1, 0.18, 0.21 \AA$^{-1}$, respectively. For the elliptical electron pocket ($\alpha$), the $k_\mathrm{F}$ values of the semi-minor and semi-major axis are estimated as 0.105 and 0.36 \AA$^{-1}$, respectively. All these values align well with those determined from quantum oscillation measurements \cite{ScSb_PRB_2018}. By approximating the spherical shape of the hole pockets and elliptical shape of electron pockets, we have estimated the hole and electron concentration to be $\sim$ $5.3\times10^{20}$ cm$^{-3}$ and $\sim$ $4.0\times10^{20}$ cm$^{-3}$, respectively. This gives $n_h\slash n_e$ ratio of $\sim$ 1.3. Overall, these values are in close agreement with our transport results.

\section{Conclusions}
In summary, the topologically trivial semimetal ScSb is found to exhibit a large Nernst thermopower of 47 $\mu$V/K at 14 T and 12 K, resulting in a high Nernst power factor of 35 $\times 10^{-4}$ W m$^{-1}$ and an excellent Nernst figure of merit of 28 $\times 10^{-4}$ K$^{-1}$. These impressive values, peaking at low temperatures around 12K, arise from well compensated electron and hole carriers within 5$\%$ uncertainty and a strong phonon drag effect. It might be possible to further enhance the Nernst effect in ScSb samples with a finely tuned the electron-hole composition. Interestingly, the transport features observed in trivial ScSb closely resemble those seen in topological semimetals. Given that ScSb is topologically trivial, our findings suggest that the anomalous Nernst effect driven by Berry curvature in topological semimetals is not the primary contributor to their enhanced thermomagnetic performance. With its cubic crystal structure, ScSb is expected to perform well in both single-crystalline and polycrystalline forms. By reducing electron scattering at grain boundaries while promoting boundary phonon scattering, polycrystalline ScSb shows strong potential for use in thermomagnetic refrigeration applications.

\section*{ACKNOWLEDGMENTS}
The research at Brookhaven National Laboratory was supported by the U.S. Department of Energy, Office of Basic Energy Sciences, Contract No. DE-SC0012704. ARPES measurements used resources at 21-ID (ESM) beamline of the National Synchrotron Light Source II, a U.S. Department of Energy (DOE) Office of Science User Facility operated for the DOE Office of Science by Brookhaven National Laboratory under Contract No. DE-SC0012704.

%

\end{document}